\documentclass[twocolumn]{aastex631}
\shorttitle{Cosmic-ray production at pc-scale blobs in NGC~1068}
\shortauthors{Michiyama et al.}
\graphicspath{{./}{figures/}}

\begin{document}

\title{ALMA detection of parsec-scale blobs at the head of kiloparsec-scale jet in the nearby Seyfert galaxy NGC~1068}

\correspondingauthor{Tomonari Michiyama}
\email{t.michiyama.astr@gmail.com}

\author[0000-0003-2475-7983]{Tomonari Michiyama}
\affiliation{Department of Earth and Space Science, Graduate School of Science, Osaka University, 1-1, Machikaneyama, Toyonaka, Osaka 560-0043, Japan}
\affiliation{National Astronomical Observatory of Japan, National Institutes of Natural Sciences, 2-21-1 Osawa, Mitaka, Tokyo, 181-8588}

\author[0000-0002-7272-1136]{Yoshiyuki Inoue}
\affiliation{Department of Earth and Space Science, Graduate School of Science, Osaka University, 1-1, Machikaneyama, Toyonaka, Osaka 560-0043, Japan}
\affiliation{Interdisciplinary Theoretical \& Mathematical Science Program (iTHEMS), RIKEN, 2-1 Hirosawa, Saitama, 351-0198, Japan}
\affiliation{Kavli Institute for the Physics and Mathematics of the Universe (WPI), UTIAS, The University of Tokyo, 5-1-5 Kashiwanoha, Kashiwa, Chiba 277-8583, Japan}

\author{Akihiro Doi}
\affiliation{The Institute of Space and Astronautical Science, Japan Aerospace Exploration Agency, 3-1-1 Yoshinodai, Chuou-ku, Sagamihara, Kanagawa 252-5210,
Japan}
\affiliation{Department of Space and Astronautical Science, SOKENDAI, 3-1-1 Yoshinodai, Chuou-ku, Sagamihara, Kanagawa 252-5210, Japan}

\author[0000-0002-7576-7869]{Dmitry Khangulyan}
\affiliation{Graduate School of Artificial Intelligence and Science, Rikkyo University, Nishi-Ikebukuro 3-34-1, Toshima-ku, Tokyo 171-8501, Japan}








\begin{abstract}
We present Atacama Large Millimeter/submillimeter Array observations at $\approx100$~GHz with $0\farcs05$ (3\,pc) resolution of the kiloparsec-scale jet seen in the nearby Seyfert galaxy NGC~1068, and we report the presence of parsec-scale blobs at the head of the jet. The combination of the detected radio flux ($\approx0.8$\,mJy), spectral index ($\approx0.5$), and the blob size ($\approx10$\,pc) suggests a strong  magnetic field of $B\approx240\,\mu$G. Such a strong magnetic field most likely implies magnetic field amplification by streaming cosmic rays.
The estimated cosmic-ray power by the jet may exceed the limit set by the star formation activity in this galaxy. This result suggests that even modest-power jets can increase the galactic cosmic-ray content while propagating through the galactic bulge.
\end{abstract}



\section{Introduction} \label{sec:intro}
Cosmic rays are ultrarelativistic particles that form an important component of the cosmic background.
The cosmic-ray energy spectrum suggests local sources that are capable of boosting the particle energy beyond 1~PeV \citep{Blasi_2013A&ARv..21...70B,Kotera2011ARA&A..49..119K}.
In the Milky Way, supernovae explosions give rise to sufficient cosmic-ray sources for supplying the required accelerating power.
The cosmic-ray level in some of the observed galaxies is consistent with that of the Milky Way.
For example, the detected gamma-ray flux in nearby starburst galaxies, such as NGC~253 and M~82, is reasonably explained by starburst activities \citep{Persic_2008A&A...486..143P, Rephaeli_2010MNRAS.401..473R, Abdo_2010ApJ...725L..73A}.
However, several nearby galaxies exhibit an excess gamma-ray flux above the calorimetric limit of their star formation activity \citep{Eichmann_2016ApJ...821...87E, Ajello_2020ApJ...894...88A}.
Understanding the dominant cosmic-ray sources other than starburst activities in galaxies are urgent topic in high energy astrophysics.
Unveiling the cosmic-ray production activities in galaxies is also important for studying the galaxy evolution, as shown in recent cosmological simulations \citep{Hopkins2020MNRAS.492.3465H,Hopkins_2021MNRAS.501.4184H}.

The nearby Seyfer-2 galaxy NGC~1068 located at a distance of $D_{\rm L}=13.97\pm2.1$~Mpc 
(\citealt{Anand2021MNRAS.501.3621A})
is one of the brightest gamma-ray emitters among nonblazar galaxies 
\citep{Ackermann_2012ApJ...755..164A} 
and its starburst activity falls below the detected gamma-ray flux level \citep{Lenain_2010A&A...524A..72L, Eichmann_2016ApJ...821...87E}.
In addition, hints of the high-energy neutrino from the direction of NGC~1068 are reported \citep{Aartsen_2020PhRvL.124e1103A}.
Therefore, this galaxy is an ideal target for investigating alternative cosmic-ray sources other than those driven by the star formation activity.

In centimeter radio continuum, NGC~1068 has a prominent linear structure with an extent of 13$^{\prime\prime}$, which is confirmed by Very Large Array (VLA) \citep{Wilson_1987ApJ...319..105W}. 
This structure is considered to be a kiloparsec-scale jet. 
The distance from the central black hole to the head of the jet is $l_{\rm jet}\approx670$~pc, assuming that the jet is inclined to the line of sight by 45$^\circ$.
\citet{Garc_2014A&A...567A.125G} estimated a jet power of $P_{\rm jet}=1.8\times{10}^{43}$~erg~s$^{-1}$ based on the 1.4\,GHz map \citep{Gallimore1996ApJ...458..136G_table} assuming the phenomenological relation of 1.4~GHz luminosity and jet power \citep{Brzan_2008ApJ...686..859B}.
We apply a self-similar fluid model developed for extragalactic jet sources \citep{Kaiser_1997MNRAS.286..215K, Gallo_2005Natur.436..819G}, where the jet supplies energy at a constant rate and expands with a velocity of $v_{\rm exp}$ in a medium of constant mass density $\mu\bar{n}$
($\mu$ is the average particle mass of 0.68$m_{\rm H}$, $m_{\rm H}$ is the hydrogen mass, and $\bar{n}=1$\,cm$^{-3}$ is the average particle density).
Subsequently, the jet age is given as $t_{\rm jet}={(l_{\rm jet}}/2)^{5/3}{(\mu\bar{n}/P_{\rm jet})}^{1/3}$ by balancing the interior pressure and ram pressure of the shocked interstellar medium (ISM).
By differentiating both sides of the equation with respect to time, the lobe expansion velocity can be obtained as $v_{\rm exp}=(3/5)(l_{\rm jet}/t_{\rm jet})$; 
for NGC~1068, we obtained $t_{\rm jet}\approx1\times{10}^5$~yr and $v_{\rm exp}\approx3\times10^3$~km~s$^{-1}$.
We note that the density assumption is not very important in estimating the jet properties because the $t_{\rm jet}$ and $v_{\rm exp}$ timescales  weakly depend on the density (e.g., $v_\mathrm{exp}\propto\bar{n}^{-1/3}$).
In this Letter, we explain the cosmic-ray production activities at parsec-scale blobs based on ALMA high resolution map of the NGC~1068 jet head.

\section{Observations} \label{sec:observation}
\begin{figure*}[ht!]
\epsscale{1.1}
\plotone{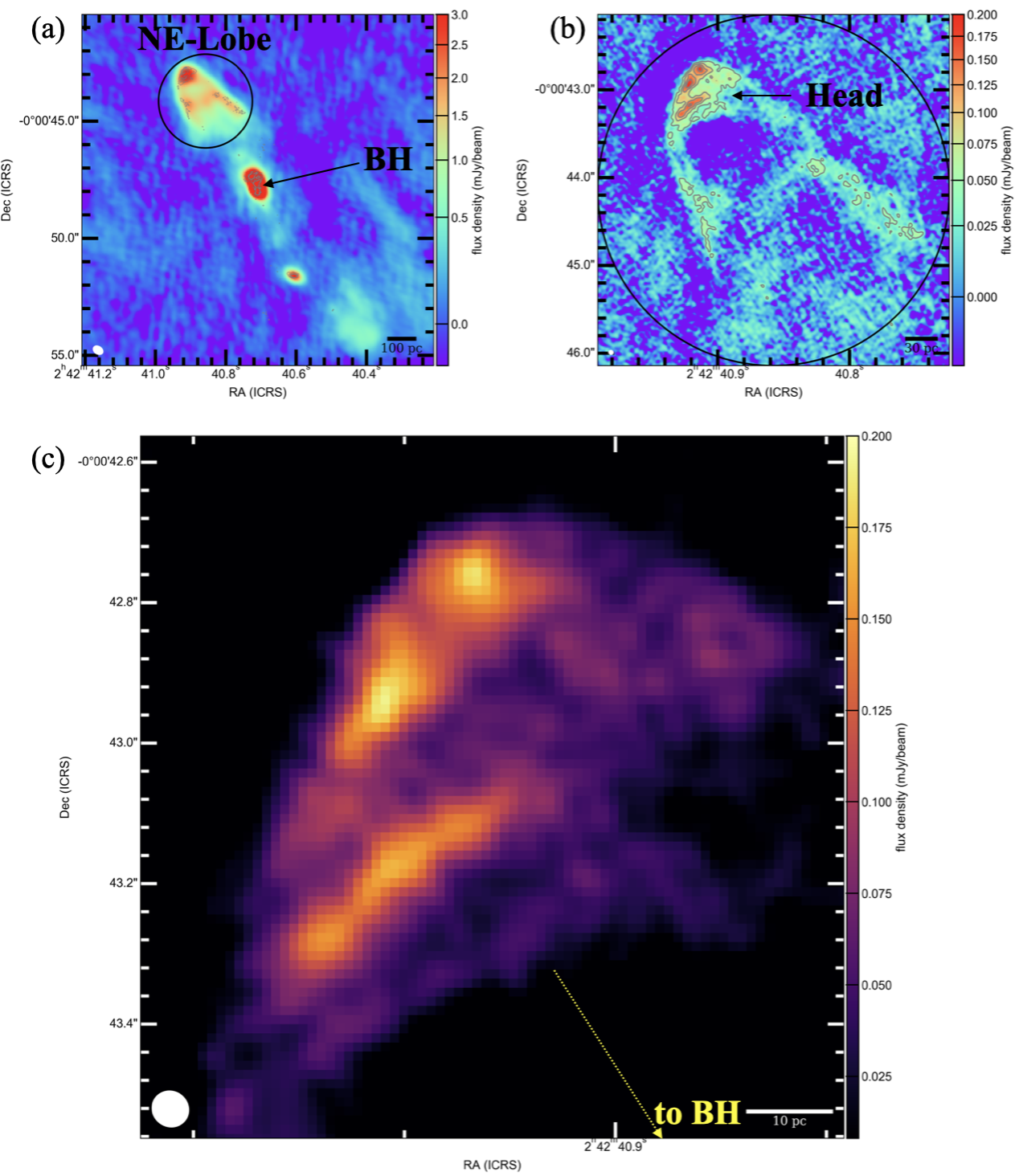}
\caption{(a) ALMA low-resolution (synthesized beam is $0\farcs46\times0\farcs34$) 93~GHz map. The solid black circle indicates the 4$\farcs$ aperture. (b) ALMA high-resolution (synthesized beam is  $0\farcs053\times0\farcs050$) 92~GHz map. To highlight the image noise, the colour was scaled by a square-root stretch. (c) Enlarged views around the head of the NE-Lobe. The dashed yellow arrow indicates the direction of the central SMBH. To highlight the structures, the colour was scaled linearly. For (a) and (b), the grey contour level is (5, 10, 15, 20) $\times\sigma$, where $\sigma = 8~\rm{\mu Jy beam^{-1}}$ is the noise level of the high-resolution map. For (a)-(c), the white ellipse at the bottom left corner represents the synthesized beam, and the black and white bars at the bottom right corner represent the physical scale bar. \label{fig:ALMA_B3}}
\end{figure*}

\begin{figure*}[ht!]
\epsscale{1.2}
\plotone{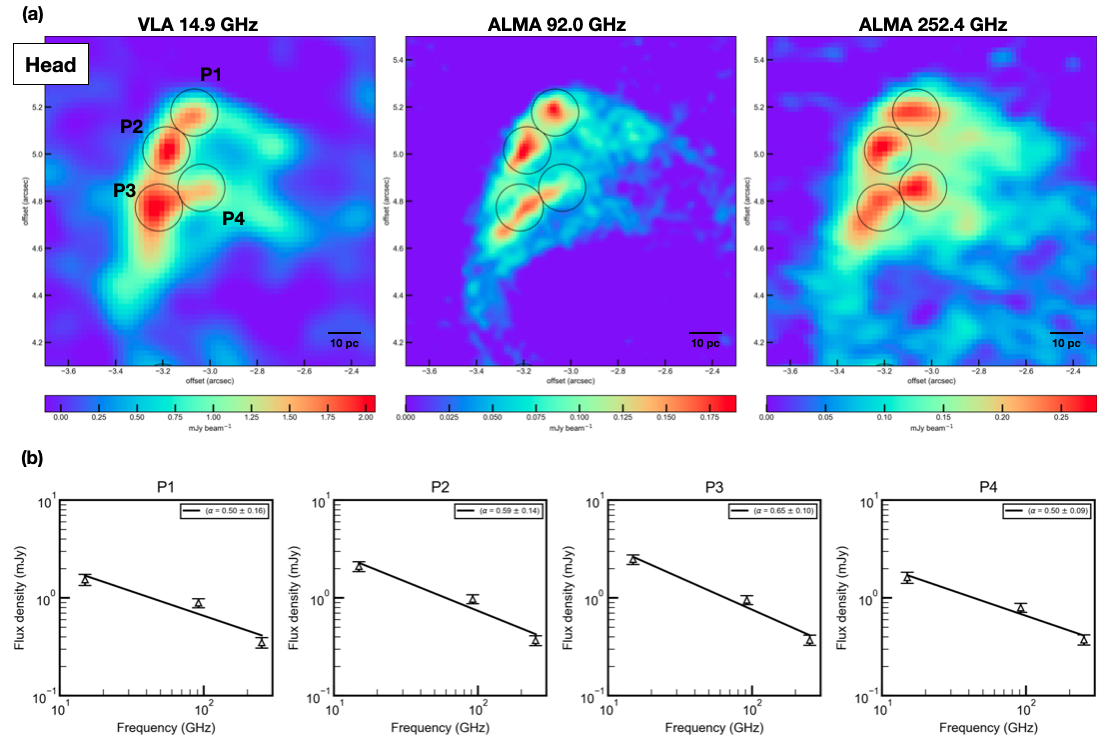}
\caption{(a) High-resolution VLA and ALMA images around the head of the NE-Lobe. The 0\farcs2 aperture around the peaks are shown as black circles (P1, P2, P3 and P4). The central coordinate for each peak is (RA, DEC) = ($02^{\rm{h}} 42^{\rm{m}} 40\fs9135$, $-00\degr 00\arcmin 42\farcs7683$) for P1, 
(RA, DEC) = ($02^{\rm{h}} 42^{\rm{m}} 40\fs9214$, $-00\degr 00\arcmin 42\farcs9287$) for P2, 
(RA, DEC) = ($02^{\rm{h}} 42^{\rm{m}} 40\fs9235$, $-00\degr 00\arcmin 43\farcs1716$) for P3, and 
(RA, DEC) = ($02^{\rm{h}} 42^{\rm{m}} 40\fs9115$, $-00\degr 00\arcmin 43\farcs0876$) for P4. The coordinates are determined in the ALMA images, and the flux for the 14.9-GHz VLA image is measured based on the offset from the nucleus.
(b) Spectral index measurements. The flux measurement apertures of each region are shown in the panel (a).
For conservative analysis, we applied the systematic $10\%$ errors.
\label{fig:index}}
\end{figure*}

Figure~\ref{fig:ALMA_B3} shows the low ($\sim0\farcs4$, 30~pc) and high ($\sim0\farcs05$, 3~pc) resolution millimeter maps of NGC~1068 by ALMA.
The low-resolution map at 93.5~GHz (Figure~\ref{fig:ALMA_B3}a) displays a kiloparsec-scale radio lobe (hereafter NE-Lobe), indicating the shock formed by the interaction of the jet with the ISM at the edge of NE-Lobe, which is consistent with the centimeter images obtained by VLA \citep{Wilson_1987ApJ...319..105W}. The brightest radio emission comes from the head of the NE-Lobe. Figure~\ref{fig:ALMA_B3}b and \ref{fig:ALMA_B3}c display the high-resolution 92~GHz map of the entire NE-Lobe region and the enlarged view of the head of the NE-Lobe, respectively.
This ALMA high-resolution map shows the bright region at the head of the NE-Lobe resolved into several blobs.
To investigate the spectral index ($\alpha\equiv-{\rm d}S_\nu/{\rm d}\nu$) of the blobs at the centimeter/millimeter, we use archival 15\,GHz VLA and 252\,GHz ALMA maps (Figure~\ref{fig:index}a) and we identified four blobs (P1--P4).
We measured the flux densities of the blobs using the {\tt imstat} command in Common Astronomy Software Applications package ({\tt CASA}) \citep{McMullin_2007ASPC..376..127M}.
When measuring flux, we smoothed the VLA and ALMA images into 0\farcs15 beam using the {\tt imsmooth} task in {\tt CASA} to reduce systematic errors due to beam dilution.
Although there are uncertainties related to the missing flux that impede the precise determination of the spectral index, our measurements show that $\alpha\approx0.5$ is preferred to hard ($\alpha>1$) or flat ($\alpha\approx0$) indices at the parsec-scale blobs.
The maximum recovery scale (MRS) of the 92.0\,GHz ALMA band ~3 data was lower than that of the 14.9\,GHz VLA and 252.4~GHz ALMA band 6 data. 
The missing flux owing to mismatched (u,v) coverage has likely no effect on the 92\,GHz flux density because the 92\,GHz flux is higher than a simple power-law model joining 15 GHz to 250 GHz (Figure~\ref{fig:index}b).
We used the archival FITS image files (not visibility) obtained from the NRAO VLA Archive Survey and the Japanese Virtual Obsefrvatory (JVO) for all the measurements in Figure~\ref{fig:ALMA_B3} and \ref{fig:index}. Table~\ref{tab:VLA_ALMA} shows the detailed information of the archival FITS image files.

\begin{figure*}[ht!]
\epsscale{0.7}
\plotone{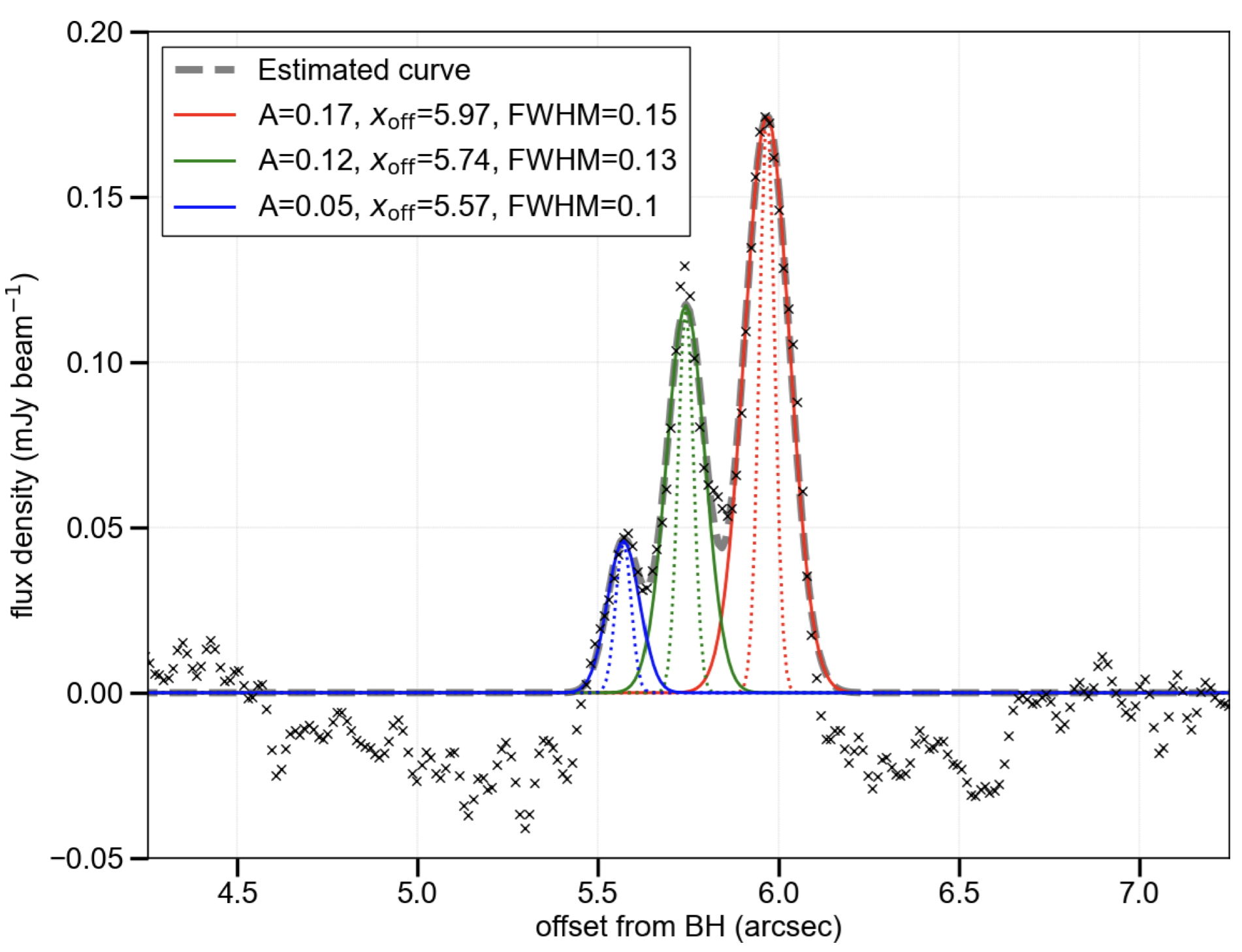}
\caption{One-dimensional 92~GHz brightness profile along the extended yellow dashed arrow in Figure 1. The profile (black crosses) was fitted by three Gaussian distribution functions (red, green, and blue solid lines). A, $x_{\rm off}$, and FWHM in the graph legends correspond to the amplitude (Jy~beam$^{-1}$), peak position ($\farcs$), and emission width ($\farcs$) obtained from the FWHM of the best fit Gaussians peak, respectively. The grey dashed line indicates the summation of the three Gaussian profiles. The dotted red, green, and blue lines indicate the three sources with the FWHM of 0$\farcs$05 (i.e., synthesized beam), suggesting that the structure along the jet direction is spatially resolved.\label{fig:profile}}
\end{figure*}

The spatial intensity profile from the central black hole to the peak position at the head is presented in Figure~\ref{fig:profile}. The blob diameter is defined as the full width at half maximum (FWHM) of the spatial profile. 
For the brightest blob, we obtained a diameter of $d_{\rm b}\approx10$~pc, which exceeds the beam size by a factor of three. 
We used the 1-D Slice tool in the viewer implemented in {\tt CASA} to make Figure~\ref{fig:profile}. The direction of the line is from the peak pixel\footnote{This position is often labeled as S1 in literature (e.g., \citealt{Gallimore_1996ApJ...464..198G_Cindex}).} in 92\,GHz map ($\alpha_{\rm ICRS}$, $\delta_{\rm ICRS}$)= (02h42m40.709s, -00d00m47.945s) to P2. The profile was fitted with three Gaussian functions using the {\tt curve\_fit} task in the Python SciPy module \citep{SciPy2020NatMe..17..261V}. The error of the best-fit value was estimated as the $95~\%$ confidence interval; for example, $A=0.175\pm0.006$~mJy, $x_{\rm off}=0\farcs967\pm0\farcs003$, and FWHM=$0\farcs1472\pm0\farcs003$ for the red curve. In this case, the most important parameter is FWHM, which is approximately three times larger than the synthesized beam size of $\approx0\farcs05$.
The 92~GHz flux density associated with the brightest blob is $\approx0.77$~mJy (per 0\farcs15 aperture in diameter), 
corresponding to a specific luminosity of $L_{\rm 92GHz}\approx1.8\times{10}^{19}{\rm\ W\ {\rm Hz}^{-1}}$.

\section{Discussion} \label{sec:discussion}
Based on ALMA results, we investigate the possible cosmic-ray production activities by the kiloparsec-scale jet in NGC~1068. We confirm the enhancement of the magnetic filed at the blobs and investigate the possible power available for the cosmic-ray acceleration as described below.

\subsection{Magnetic Field of The Blobs}\label{sec:B}
Unless the plasma density is unfeasibly high, \(>10^3\rm\,cm^{-3}\), the detected radio emission has synchrotron origin, i.e., produced by relativistic electrons interacting with magnetic field.
Assuming a power-law electron energy distribution, $n(E){\rm d}E\propto E^{-p}{\rm d}E$, the electron power-law index of $p\approx2$ can be anticipated from the obtained radio spectra (note that $p=2\alpha+1$ where $\alpha=0.5$).
The most commonly used approach is ``minimum-energy formula".
This approach requires that the total energy density of the magnetic field ($U_{\rm B}$) and the radio-emitting electrons ($U_{\rm e}$) is close to the minimum value.
According to the formula based on equation (16.43) and (16.44) of \citet{longair_2011}\footnote{This formulation corresponds equation (1) of \citet{Beck2005AN....326..414B}, represented as $B_{\rm class}$.},
\begin{eqnarray}\label{eq:B_SI}
B_\mathrm{min}&=&\left[ \frac{3\mu_0}{2}\frac{G(\alpha) L_{\rm \nu}}{V} \right]^{\frac{2}{7}}\\ \nonumber
&\approx& 240\ {\rm \mu G}\\ \nonumber
&&\times\left(\frac{L_{\rm 92GHz}}{1.8\times{10}^{19}\ {\rm W\ Hz}^{-1}}\right)^\frac{2}{7}\left(\frac{V}{{5.4\times10}^{2}\ {\rm pc^3}}\right)^{-\frac{2}{7}},
\end{eqnarray}
where
\begin{eqnarray}
G(\alpha) &=& \frac{1}{a(p)(p-2)}[\nu_{\rm min}^{-(p-2)/2}-\nu_{\rm max}^{-(p-2)/2}]\nu^{(p-1)/2}\\ \nonumber
&&\times \frac{(7.4126\times10^{-19})^{-(p-2)}}{2.344\times10^{-25}}(1.253\times10^{37})^{-(p-1)/2},
\end{eqnarray}
$\nu_{\min}=15$~GHz, 
$\nu_{\max}=250$~GHz, 
$\nu=92$~GHz,
$\mu_0=1.3\times10^{-6}$~m~kg$^{-2}$~A$^{-2}$,
and 
$a(2)=0.529$.
The $B_{\rm min}\approx240\,{\rm \mu G}$ corresponds $U_{\rm B}=B^2/(8\pi)=2.2\times10^{-9}\,{\rm erg\,cm^{-3}}$ and $U_{\rm e}=(4/3)U_{\rm B}=3.0\times10^{-9}\,{\rm erg\,cm^{-3}}$.
This magnetic field strength is significantly higher than that of the ISM (a few $\mu$G) \citep{Ferri_2001RvMP...73.1031F}, suggesting a strongly amplified magnetic field. Such a magnetic field amplification is observed in Galactic supernova remnants, where cosmic-ray streaming enhances the ISM magnetic field up to a few hundred $\mu$G \citep[][see also \citealt{2022Sci...376...77A}]{Bell_2013MNRAS.431..415B}. Thus, we applied $B=B_{\rm min}=240$~$\mu$G as a fiducial value assuming the magnetic field amplification by the cosmic-ray instability seen in SNRs.

Locally accelerated electrons lose their energy owing to synchrotron cooling over a time scale of
\begin{eqnarray}
t_{\rm sync}&=&\frac{3}{4}\frac{m_{\rm e}c}{\sigma_{\rm T}U_{\rm B}}\gamma_{\rm e}^{-1} \\ \nonumber 
&\approx&{4.5\times10^4}\ {\rm yr}\ \left(\frac{B}{240\ {\rm \mu G}}\right)^{-\frac{3}{2}}\left(\frac{\nu_{\rm sync}}{\rm 92\ GHz}\right)^{-\frac{1}{2}}.
\end{eqnarray}
Here, $\nu_{\rm sync}=3eB\gamma_{\rm e}^2/4\pi m_{\rm e}c$, where $m_{\rm e}$ is the electron rest mass, $\sigma_{\rm T}$ is the Thomson scattering cross-section, $U_{\rm B}=B^2/(8\pi)$ is the magnetic field energy density, and $e$ is the elementary charge.
For comparison, the advection timescale through the blob is
$t_{\rm adv} = d_{\rm b}/v_{\rm d}\approx1\times{10}^4$~yr
(adopting downstream speed, $v_{\rm d} = v_{\rm exp}/4$, expected for strong shocks).
Given that $t_{\rm adv} \lesssim t_{\rm sync}$, synchrotron emission is produced in the slow-cooling regime, implying that the measured electron spectrum is produced directely by the acceleration process.

\subsection{Maximum Cosmic-ray Energy in the Kiloparsec-scale Jet}\label{Emax}
The measured radio spectrum with $\alpha=0.5$ implies a power-law electron spectrum with $p\approx2$, which is consistent with the canonical slope predicted for diffusive shock acceleration under a strong shock \citep{Blandford_1978ApJ...221L..29B, Bell_1978MNRAS.182..147B}.
The bright synchrotron emission of non-thermal electrons at the shock also implies efficient acceleration of the protons.
If a significant upstream current is generated by cosmic-ray particles, the required amplification of the magnetic field is possible by nonresonant hybrid instability \citep{Bell_2013MNRAS.431..415B}. This implies that a considerable fraction of the downstream energy is transferred to relativistic protons.
Because the physical conditions revealed at the forward shock are similar to those at the blast wave produced by a supernovae explosion, we can readily use the estimates for the cosmic-ray maximum energy from the literature.
The maximum cosmic-ray energy can be calculated using Equation 6 in \citet{Bell_2013MNRAS.431..415B} as
\begin{eqnarray}
E_{\rm cr,max} &=& 8~{\rm PeV}
\left(\frac{n_{\rm e}}{1\ {\rm cm}^{-3}}\right)^{\frac{1}{2}}\left(\frac{v_{\rm shock}}{3\times10^3~{\rm km\ s^{-1}}}\right)^3 \nonumber\\
&&\times\left(\frac{t_{\rm age}}{{1\times10^5~\rm yr}}\right)
\times\left(\frac{p_{\rm cr}/\rho v_{\rm shock}^2}{0.3}\right),
\end{eqnarray}
where $n_{\rm e}$ is the electron density, $v_{\rm shock}=v_{\rm exp}$ is the forward shock velocity, $t_{\rm age}$ is the age of the system, and 
$p_{\rm cr}/\rho v_{\rm shock}^2$ is the cosmic-ray pressure ratio at the shock.
For NGC 1068, we apply 
$n_{\rm e}=1~{\rm cm}^{-3}$ (as a typical ISM value), 
$v_{\rm shock}=3\times10^3~{\rm km\ s^{-1}}$, 
$t_{\rm age}=t_{\rm jet}=1\times{10}^5$~yr, 
and $p_{\rm cr}/\rho v_{\rm shock}^2=0.3$ (assuming the same situation as in the supernova cases).
The maximum cosmic-ray energy can also be predicted on the basis of the requirement that ``the cosmic-ray acceleration timescale ($t_{\rm accel}$) cannot exceed the age of the system".
Using the common assumption of Bohm diffusion, the cosmic-ray acceleration timescale was estimated to be $t_{\rm accel}\approx(10\eta_{\rm B}cr_{\rm g})/(3v_{\rm shock}^2)$, where $r_{\rm g}=E_{\rm cr}/eB$ is the gyroradius and $\eta_{\rm B}$ is the Bohm factor.
When we apply $t_{\rm adv}$ (the shortest time among $t_{\rm adv}$, $t_{\rm sync}$, and $t_{\rm jet}$) as the age of the system, the maximum cosmic-ray energy is given by
\begin{eqnarray}\label{eq:Emax}
E_{\rm cr,max} &\approx& \frac{3e}{10c} \eta_B^{-1} B v_{\rm shock}^2 t_{\rm adv} \\
&\approx& 30\,\eta_B^{-1}\ {\rm PeV} \nonumber \\
&&\times\left(\frac{B}{230\ {\rm \mu G}}\right)\left(\frac{v_{\rm shock}}{3\times10^3\ {\rm km\ s^{-1}}}\right)^2\left(\frac{t_{\rm adv}}{{1\times10^4~\rm yr}}\right), \nonumber
\end{eqnarray}
Both these estimates yield cosmic-ray maximum energies of $\approx$PeV.

The reverse shock at the jet head region (not the forward shock propagates through the ISM medium) might be a possible acceleration site \citep{Araudo2016MNRAS.460.3554A}. However, the reverse shock is unlikely the origin of the jet head in NGC~1068 based on the energy argument. The average magnetic field of $B_{\rm ave}\approx50$\,$\mu$G at the jet head (i.e., the regions shown in Figure~\ref{fig:ALMA_B3}c) can be obtained from the total flux of $\approx9$\,mJy and the volume a hemisphere of radius $r_{t}\approx40$\,pc).
Assuming that this structure is produced by the reverse shock, the ratio between the energy density of the magnetic field $B_{\rm ave}^2/(8\pi)$ and the energy density of the jet $P_{\rm jet}/(\pi r_{\rm t}^2 c)$ is calculated as $\approx0.008$, which means that only $<1\%$ of the power of the relativistic jet is converted into the magnetic energy in the reverse shock, which is not likely the common case \citep{Blandford1977MNRAS.179..433B}.

\subsection{Cosmic-ray Power of the Kiloparsec-scale Jet}
The VLA and ALMA data allow us to obtain the total electron power as
$P_{\rm e}\approx\pi{(d_{\rm b}/2)}^2v_{\rm d}U_{\rm e}\approx 2\times{10}^{38}$~erg~s$^{-1}$ per a radio-emitting blob. Here, we use the energy density $U_\mathrm{e}$ derived in Section~\ref{sec:B}.
This estimate for $P_{\rm e}$ implies that the total cosmic-ray accelerating power in NGC~1068 can be as large as $P_{\rm CR,blobs}\approx1\times{10}^{41}$~erg~s$^{-1}$, where we summed the contributions from the four detected blobs, accounted for the existence of the counter jet, and adopted  $K_{\rm ep}\approx0.01$. The expected power available for the cosmic-ray acceleration is $\approx1\%$ of the jet power, which is comparable to the efficiency of cosmic-ray acceleration produced by supernovae remnants.
This value of $P_{\rm CR,blobs}$ corresponds to the lower limit of the total cosmic-ray accelerating power of this jet ($P_{\rm CR,jet}$), because we neglect the cosmic-ray acceleration occurring outside the blobs resolved by ALMA, i.e. $P_{\rm CR,jet}>P_{\rm CR,blobs}$.
The total cosmic-ray accelerating power of the supernovae in NGC~1068 can be estimated as $P_{\rm CR,SN}\approx2\times{10}^{41}$~erg~s$^{-1}$ where the observed supernovae rate is 0.07~per year \citep{Storchi-Bergmann_2012ApJ...755...87S, Eichmann_2016ApJ...821...87E}, the energy of a supernova is ${10}^{51}$~erg, and $10~\%$ of supernova energy is transferred to the cosmic rays.
The relation of $P_{\rm CR,jet}>P_{\rm CR,blobs} \approx P_{\rm CR,SN}$ indicates that the resolved parsec-scale blobs in the termination region of the kiloparsec-scale jet can be powerful sources of cosmic-ray production activities, as well as the star formation in NGC~1068 and would contribute to the gamma-ray excess seen in this galaxy  (if $K_{\rm ep}$ in the blobs are same as that of supernovae remnants).
Considering the spectral shape, maximum energy, and total energy budget, the blobs at the kiloparsec-scale jet head are likely important cosmic-ray factories. 

Finally, we note that there are various radio bright regions in NGC~1068 other than the NE-lobe region such as the components S1 and C, labeled by \citet{Gallimore1996ApJ...458..136G_table}. These regions would be alternative source of cosmic-ray acceleration sites and gamma-ray production regions. Regarding the component S1, it has been argued an efficient cosmic-ray accelerator, however, resulting gamma-ray emission would be strongly suppressed by the pair-creation processes because of intense photon field provided by the nucleus \citep[see e.g.,][]{Inoue2020ApJ...891L..33I}. Therefore, the component S1 can not be the major gamma-ray production site. The component C would be the other remaining candidate. However, a large radio flux of the component C alone cannot be considered strong evidence for cosmic-ray acceleration. One needs several ingredients for claiming CR acceleration: (1) operation of some acceleration mechanism; (2) presence of protons (ions) in the accelerator; (3) sufficiently high acceleration rate to boost particle energy to the PeV regime; and (4) conditions for particle escape from the acceleration site. The forward shock of the jet in NGC~1068 satisfies all these requirements, and this lucky combination allows us to claim its contribution to the CR budget. Furthermore, we see hints of a strong magnetic field amplification, which under the expected condition implies a significant current of cosmic ray, i.e., an indirect (but strong) sign of cosmic-ray acceleration. It is not clear if any of the above requirements are fulfilled in the component C. Analysis of their potential requires a reliable physical model. For P1-P4 we argue that these blobs belong to the downstream region of the jet forward shock, which allows us to estimate the key physical parameters there (e.g., the magnetic field amplification factor, the flow speed). Since, to our understanding, there is no physical model for the bright region C, we leave the analysis of the cosmic-ray acceleration potential of these regions beyond the scope of this letter.

\section{Summary}
Supernovae are considered to be the dominant source of cosmic-ray production in galaxies.
However, recent gamma-ray observations revealed galaxies whose cosmic-ray power is beyond the calorimetric limit of the star formation activities. Because cosmic rays and their feedback processes play a crucial role in the evolution of galaxies, the identification of the cosmic-ray factories in external galaxies is intriguing.
We show that the kiloparsec-scale jets observed in one such limit-break galaxy; NGC~1068, is a powerful cosmic-ray production site based on high-resolution (0\farcs05) ALMA maps of the termination shock region.
The radio spectrum showed a spectral index of $\approx0.5$, corresponding to the electron spectral index of $\approx2$, which is consistent with the canonical value predicted by the diffusive shock acceleration.
The amplified magnetic field is necessary to explain the radio flux at the blobs and the level of amplification is consistent with the cosmic-ray streaming instability that occurs in Galactic supernova.
The maximum cosmic-ray energy may achieve $\approx$PeV and the cosmic-ray power at this kiloparsec-scale jet may be greater than that estimated based on the supernovae rate in NGC~1068.
These results suggest that cosmic-rays can be generated far from the central black hole owing to interactions between the kpc-scale jet and ISM. 
Considering kiloparsec-scale jet as a new dominant cosmic-ray accelerator in a galaxy is important in comprehending the impact of cosmic rays on the evolution of galaxies.

\begin{acknowledgments}
T.M. and Y.I. appreciate support from NAOJ ALMA Scientific Research Grant Number 2021-17A.
T.M. is supported by JSPS KAKENHI grant No. 22K14073. 
DK acknowledges support by JSPS KAKENHI grants No. 18H03722, 18H05463, and 20H00153.
Some of the ALMA data were retrieved from the JVO portal (\url{http://jvo.nao.ac.jp/portal/}) operated by ADC/NAOJ. Data analysis was in part carried out on the common use data analysis computer system at the Astronomy Data Center, ADC, of the National Astronomical Observatory of Japan. This letter makes use of the following ALMA data: $\#$2018.1.01135, $\#$2018.1.01506, $\#$2015.1.01144, $\#$2016.1.00232, $\#$2016.1.00023. ALMA is a partnership of ESO (representing its member states), NSF (USA), and NINS (Japan), together with NRC (Canada), NSC, ASIAA (Taiwan), and KASI (Republic of Korea), in cooperation with the Republic of Chile. The Joint ALMA Observatory is operated by ESO, AUI/NRAO, and NAOJ.

%
\vspace{5mm}
\facilities{ALMA, VLA}
\software{astropy \citep{2013A&A...558A..33A,2018AJ....156..123A},
ALMA Calibration Pipeline,
CASA \citep{McMullin_2007ASPC..376..127M}
}
\end{acknowledgments}

%

\begin{deluxetable*}{lcccccccc}
\tablenum{1}
\tablecaption{Archival FITS images\label{tab:VLA_ALMA}}
\tablewidth{0pt}
\tablehead{
\colhead{telescope} & \colhead{freq.} & \colhead{resolution} & \colhead{rms} & \colhead{date} & \colhead{band} & \colhead{config.} & \colhead{MRS} & \colhead{ALMA ID} \\
\colhead{} & \colhead{(GHz)} & \colhead{} & \colhead{($\mu$Jy~beam$^{-1}$)} & \colhead{} & \colhead{} & \colhead{} & \colhead{} & \colhead{}
}
\decimalcolnumbers
\startdata
VLA & 14.9 & $0\farcs145\times0\farcs136$ & 105 & 1983-11-03 & U & A/A & 1\farcs8 & $-$ \\
ALMA & 92.0 & $0\farcs053\times0\farcs050$ & 8 & 2019-06-22,24 & 3 & C43-9/10 & 0\farcs9 & 2018.1.01135.S \\
ALMA & 93.5 & $0\farcs46\times0\farcs34$ & 43 & 2019-09-22,23 & 3 & C43-6 & 5\farcs5 & 2018.1.01506.S \\
ALMA & 252.4 & $0\farcs150\times0\farcs103$ & 19 & 2017-07-23$^{*1}$ & 6 & C43-6$^{*2}$ & 34 & 2016.1.00023.S \\
\enddata
\tablecomments{The values of observed (sky) frequency and resolution are based on the header of the downloaded FITS files. The sensitivity was measured in the emission-free region around the NE-Lobe. The MRS for VLA observations was predicted using a table available on the VLA website. The MRS for ALMA observations was based on the observatory reports (QA2 report). *1 Observations were conducted on 2017-07-23, 2017-07-24, 2017-07-27, 2017-12-27, 2018-01-01, 2018-01-19, and 2018-01-21. *2 Observations were conducted in C40-5 and C43-5 as well.}
\end{deluxetable*}




\clearpage
\bibliography{sample631}{}
\bibliographystyle{aasjournal}

\end{document}